\documentclass[useAMS,usenatbib]{mn2e-1}%{/Users/davidnataf/MNRASLetter1/mnras/mn2e2}
\usepackage{graphicx}
\usepackage{array,longtable}
\usepackage{natbib}
\usepackage{float}
\usepackage[T1]{fontenc}
\usepackage{ae}
\usepackage{aecompl}
\usepackage{amsmath, amssymb, amsfonts}
\usepackage{multicol}

\title[Interstellar Extinction and M101 Cepheids]{Uncertainties in The Interstellar Extinction Curve and the Cepheid Distance to M101}
\author[Nataf]{David M. Nataf$^1$\thanks{Email: david.nataf@anu.edu.au}
\vspace*{6pt}\\
$^{1}$Research School of Astronomy and Astrophysics, The Australian National University, Canberra, ACT 2611, Australia  }
\begin{document}
\include{journaldefs}
\date{Accepted 2015 January 21.  Received 2014 December 26; in original form 2014 October 27}

\pagerange{\pageref{firstpage}--\pageref{lastpage}} \pubyear{2014}
\maketitle
\label{firstpage}

\begin{abstract}
I revisit the Cepheid-distance determination to the nearby spiral galaxy M101 (Pinwheel Galaxy) of \citet{2011ApJ...733..124S}, in light of several recent investigations questioning the shape of the interstellar extinction curve at $\lambda \approx 8,000$ \AA (\textit{i.e.} I-band). I find that the relatively steep extinction ratio $A_{I}/E(V-I)=1.1450$  \citep{2007ApJ...663..320F} is slightly favoured relative to  $A_{I}/E(V-I)=1.2899$  \citep{1999PASP..111...63F} and significantly favoured relative the historically canonical value of $A_{I}/E(V-I)=1.4695$  \citep{1989ApJ...345..245C}. The steeper extinction curves, with lower values of $A_{I}/E(V-I)$, yield fits with reduced scatter, metallicity-dependences to the dereddened Cepheid luminosities that are closer to values inferred in the local group, and that are less sensitive to the choice of reddening cut imposed in the sample selection. The increase in distance modulus to M101 when using the preferred extinction curve is ${\Delta}{\mu} \sim 0.06$ mag, resulting in an estimate of the distance modulus to M101 relative to the LMC of $ {\Delta}\mu_{\rm{LMC}} \approx 10.72 \pm 0.03$ (stat). The best-fit metallicity-dependence is $dM_{I}/d\rm{[O/H]} \approx (-0.38 \pm 0.14$ (stat)) mag dex$^{-1}$. \end{abstract}
\maketitle

%\citet{1989ApJ...345..245C} as the dotted-blue line, \citet{1999PASP..111...63F} as the short-dashed-green, and \citet{2007ApJ...663..320F} 

\begin{keywords}
distance scale -- stars: variables: Cepheids -- ISM: dust, extinction --  galaxies: individual: (M101) 
 \end{keywords}

\section{Introduction}
\label{sec:introduction}
Interstellar extinction causes luminosity sources to appear fainter and redder, and is thus somewhat degenerate with distance which causes luminosity sources to appear fainter but not redder.  In principle, this difference (the colour offset) can allow one to break the degeneracy between interstellar extinction and distance. If one assumes a consistent total-to-selective extinction ratio  $R_{I}   = A_{I}/E(V-I) = A_{I}/(A_{V}-A_{I}) $, one can estimate a ``Wesenheit\footnote{``Wesenheit" is the German word for ``essential".} magnitude" \citep{1982ApJ...253..575M} that is nominally reddening-independent:
\begin{equation}
W_{I} = I - R_{I} {\times} (V-I) .
\end{equation}
The values of $R_{I}$ regularly used in the literature range from 1.60 \citep{1994AJ....107.2060P}, 1.55 \citep{2011JAVSO..39..122M,2012ApJ...747...50N}, down to 1.45 \citep{2006ApJ...652.1133M,2011ApJ...743..176G,2011ApJ...733..124S}. There is certainly a strong empirical basis for this assumption.  \citet{1989ApJ...345..245C}, in their seminal investigation of extinction from the ultraviolet to the near-infrared, yielded parametric fits for the extinction curve as a function of wavelength and the free-parameter $R_{V} = A_{V}/E(B-V)$. For the case $R_{V}=3.1$, their functions yield $R_{I} \approx 1.45$. The results of \citet{1989ApJ...345..245C} are a linchpin of astronomy as a whole, with over 5,000 refereed citations at the time of writing. These values have had convincing independent support elsewhere. \citet{2003ApJ...590..284U} measured $R_{I} = 1.44 \pm 0.03$ toward the Large Magellanic Cloud (LMC), which is the anchor of the extragalactic distance scale \citep{2001ApJ...553...47F} and many investigations of Cepheids. \citet{2012ApJ...748..107P} developed a comprehensive physical model and matched to $\sim$177,000 photometric measurements from a sample of 287 LMC, Small Magellanic Cloud (SMC), and Milky Way Cepheids. They measured a best-fit mean extinction curve of $R_{V} \approx 3.127$, assuming the formalism from \citet{1989ApJ...345..245C}. Similarly, \citet{2014arXiv1409.2500P}, matched to $\sim$6,800 photometric measurements over a broad range of wavelengths, and showed that the $R_{V}=3.1$ extinction curve from \citet{1989ApJ...345..245C} is an effective fit to the $UBVRI$ light curves of Type-II plateau supernovae. 

There has been recent evidence that the extinction curve of even the diffuse interstellar medium may be variable on a systematic basis, or simply distinct than as parameterised by \citet{1989ApJ...345..245C}. \citet{2011ApJ...737..103S} studied reddening values toward Milky Way halo stars, and found that the exitnction curve of \citet{1999PASP..111...63F}, with its different method of derivation with the extinction curve fit to some cubic spline anchor points rather than seventh degree polynomials among other distinctions, and its steeper extinction in the wavelength range 6,000 \AA  $\leq \lambda \leq $9,000 \AA, was more consistent with observations that of \citet{1989ApJ...345..245C}. This finding contributed to a revision of the SFD extinction maps \citep{1998ApJ...500..525S}. Similarly, \citet{2014ASPC..482..275S} showed that reddening toward open clusters within 3 Kpc of the Sun is well-fit by an $R_{V}=3.1$ \citet{1999PASP..111...63F} extinction curve, which predicts $R_{I} \approx 1.27$ for hot stars. Further, it is by now well-documented that the inner-parts of the Milky Way, which are rigorously investigated, are fit by a steeper standard extinction curve than the  $R_{V}=3.1$  curve from  \citet{1989ApJ...345..245C}. In their study of 17,000 Galactic bulge RR Lyrae stars identified in OGLE-III photometry \citep{2011AcA....61...83S}, \citet{2012ApJ...750..169P} measure   $R_{I}=1.080 \pm 0.007$. \citet{2013ApJ...769...88N} used red clump stars (which are $\sim$2,000 Kelvin colder and $\sim10{\times}$ more metal-rich) measured with the same OGLE-III photometry and obtained  $<R_{I}>=1.218$, with significant variations confirmed spanning a range $1.0 \lesssim R_{I} \lesssim 1.4$. The large shifts in the extinction curve are observed to take place toward sightlines closer to one another than 0.5 degrees on the sky, or a transverse distance no greater than 75 parsecs. %\citet{2014A&A...564A..63M}, using \textit{Hubble Space Telescope} (HST) photometry and \textit{VLT-Flames} spectroscopy of OB stars in 30 Doradus, inferred $R_{I} \approx 1.35$.  
Finally, from the extinction study of \citet{1996AstL...22..334B}, which  based purely off photometry of Cepheids\footnote{From Table 4 of  \citet{1996AstL...22..334B} one can derive $R_{I}=0.84$, but that is because the $I$-band filter used in that study is centred at $\lambda \approx 8,800$\AA, which is at a longer wavelength than the standard Landolt filter.},  $R_{I}\approx 1.07$. 

Reading this sequence of findings some may argue that the astronomical community should simply switch to using the \citet{1999PASP..111...63F} extinction curve as default over that of \citet{1989ApJ...345..245C}. That may turn out to be the correct course of action, but as of now there at least three causes for concern with this approach. First, as discussed earlier in the introduction, the extinction curve of \citet{1989ApJ...345..245C} does have convincing independent empirical support, and this would need to be explained if the $R_{V}=3.1$ extinction curve of \citet{1989ApJ...345..245C}  turns out to be incorrect. Second, the evidence is currently overwhelming that the extinction curve of \citet{1999PASP..111...63F} fails in the near-IR, where  \citet{1999PASP..111...63F}  predicts $A_{J}/A_{K} \approx 2.31$ that is nearly independent of $R_{V}$. However, the most precisely and accurately measured value in the Galactic astronomy literature is $A_{J}/A_{K} \approx 3.02$ \citep{2009ApJ...696.1407N}. Finally, buoyed by more abundant data and a different mode of analysis, Fitzpatrick himself argued for a different mean Galactic extinction curve in \citet{2007ApJ...663..320F} (see their Table 5), which is as distinct from \citet{1999PASP..111...63F} as \citet{1999PASP..111...63F} is from  \citet{1989ApJ...345..245C}.

Separately and independently to these issues, evidence has recently emerged that the wavelength-dependence of interstellar extinction is a function of the spectral type and the specific-star-formation rate of a galaxy \citep{2013ApJ...775L..16K}, and  even location within a galaxy \citep{2009ApJ...707..510Z,2013ApJ...779...38P,2014ApJ...785..136D}. During the referee process for this paper, \citet{2014arXiv1412.2138F} found a best-fit extinction curve toward the Cepheids of NGC 4258 of $R_{V}=4.9$, which is extremely shallow,  using a combination of $UBVR$ photometry. 

Having worked extensively on some of these issues, I cannot argue at this time that there is a convincing framework unifying these disparate, seemingly mutually-inconsistent findings.  What I am confident in, is that more investigation is needed, both of the shape of the extinction curve, how it varies with direction, and how this hitherto largely ignored source of systematic error has metastasised into the general astronomy literature and its library of accepted values. There is an abundance of data available in a broad range of bandpasses for a broad range of objects observed in a broad range of contexts, and thus further research that could elucidate these issues is possible. 

Thus, I tackle the Cepheid distance to M101 for two reasons. First, as a probe of the interstellar extinction curve in and of itself. The data products of \citet{2011ApJ...733..124S} are of high quality, with the photometry obtained from the \textit{Hubble Space Telescope} (HST) and reduced via a suite of state-of-the-art software tools. The observations are of a spiral galaxy of similar mass and morphology as the Milky Way. As M101 is viewed nearly face-on, the reddening values are sufficiently small that a large number of Cepheids can be identified. Indeed, the catalogue of \citet{2011ApJ...733..124S} is one of the largest homogenously-measured extragalactic catalogues of Cepheids, containing 1,227 Cepheid candidates. Second, this investigation can help discern what impact these uncertainties might have on the field of the extragalactic distance ladder, where Cepheids are a widely-used standardisable candle \citep{2006ApJ...652.1133M,2011ApJ...730..119R,2011ApJ...743..176G,2014MNRAS.440.1138E}. As the era of precision cosmology is now over a decade old \citep{2003ApJS..148..175S}, with the reported error bars growing smaller and potential tensions growing greater \citep{2013arXiv1303.5076P}, the requirements for measurement precision are now higher-than-ever, necessitating diligent investigations of parameters such as extinction coefficients. 

\section{Defining the Problem and Methodology}
\label{sec:Methodology}
\begin{figure*}
\begin{center}
\includegraphics[totalheight=0.32\textheight]{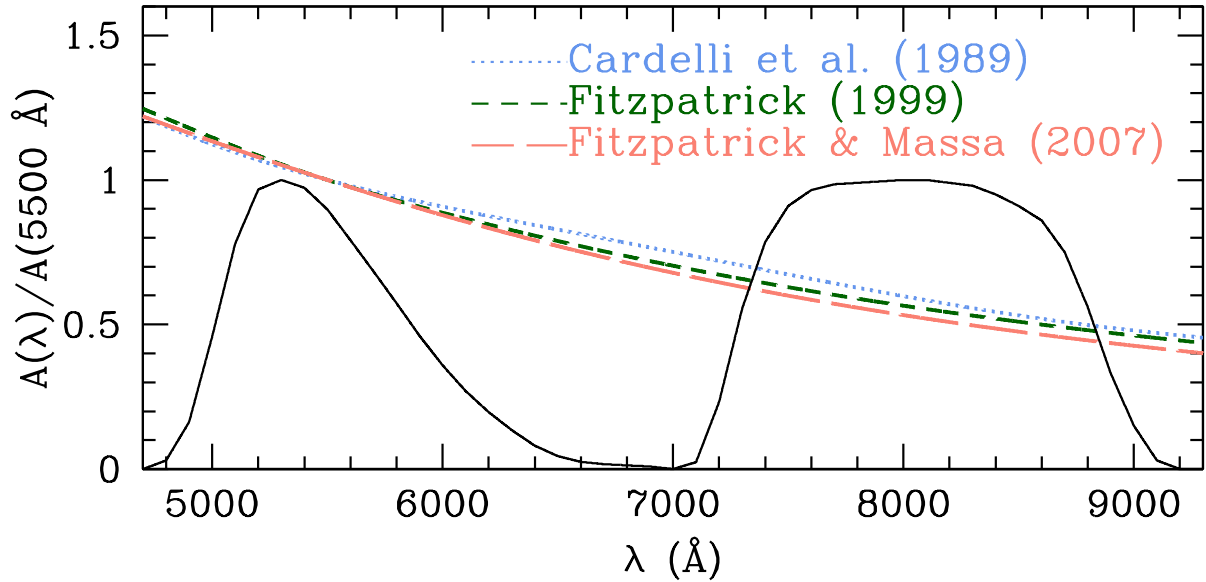}
\end{center}
\caption{\large The extinction curves of \citet{1989ApJ...345..245C} as the dotted-blue line, \citet{1999PASP..111...63F} as the short-dashed-green, and \citet{2007ApJ...663..320F} as the long-dashed-salmon line, as a function of wavelength. Also shown are the \citet{1992AJ....104..340L} $V$ and Landolt $I$ filters in black. The different trends of the extinction curves over this wavelength regime lead to different predictions for the extinction ration $A_{I}/E(V-I)$.} 
\label{Fig:ExtinctionCurves}
\end{figure*}

As discussed in the introduction, virtually any extinction coefficient in the range  $1.05 \lesssim R_{I} \lesssim 1.45$ can be assigned some empirical support ascribing it the designation of ``the standard extinction ratio for the diffuse interstellar medium." That is problematic and a source of major systematic error previously undiagnosed and unquantified in the literature on the use of Cepheids to probe the extragalactic distance ladder. In this paper, I test for three different cases. These are the extinction curves of \citet{1989ApJ...345..245C}, \citet{1999PASP..111...63F}, and \citet{2007ApJ...663..320F}, which are shown in Figure \ref{Fig:ExtinctionCurves}. I convolve these curves with the transmission functions of the \citet{1992AJ....104..340L} $V$ and $I$ filters (also shown in Figure \ref{Fig:ExtinctionCurves}), and a synthetic spectrum of a $T_{\rm{eff}}=5,750$ Kelvin, [Fe/H]$=0$, $\log{g}=1.0$ star kindly sent to me by Luca Casagrande, where the parameters are chosen to typical of Cepheids as modelled by  \citet{2012ApJ...748..107P}. The predicted extinction coefficients are listed in Table \ref{table:ExtinctionRatios}.

At the request of the referee, I also list in Table \ref{table:ExtinctionRatios2} what the extinction coefficients would be for different assumed parameters. Shifting the temperature by 1000 Kelvin only shifts the extinction coefficients by $\sim$1\% or less. The dominant uncertainty lies in the assumed extinction curve, and not the convolution of  the stellar spectrum with which the transmission filters are convolved. Nevertheless, should one actually be attempting ``1\% cosmology", then one should compute slightly different extinction coefficients for every Cepheid. With the two bandpasses used in this work, the shift in distance modulus is only $\sim$0.003 mag, with the values of Table 1 yielding the marginally smaller distances. 

\begin{table}
\caption{\large Predicted extinction coefficients for the Landolt, 2MASS, ACS, and WFC3 filter transmission curves convolved with the synthetic spectrum of a $T_{\rm{eff}}=5,750$ Kelvin, [Fe/H]$=0$, $\log{g}=1.0$ star as well as 0.40 mag of extinction at $\lambda=5500$\AA. These are the coefficients assumed by this study. } 
\large
\centering 
\begin{tabular}{|l|c|c|c|} 
\hline\hline\hline 
 % & \rot{  \citet{1989ApJ...345..245C}  }  &   \rot{\citet{1999PASP..111...63F} } &  \rot{\citet{2007ApJ...663..320F}}    \\  
    & CCM89  &  F99 &  FM07    \\  
 \hline\hline
$A_{I}/E(V-I)$ & 1.4695 & 1.2899 & 1.1450 \\
$A_{U}/A_{5500 \text{\AA}}$ & 1.5492 & 1.5532  & 1.5276 \\
$A_{B}/A_{5500 \text{\AA}}$ & 1.2893  & 1.3022 & 1.2774  \\
$A_{V}/A_{5500 \text{\AA}}$ & 1.0020  & 1.0015  & 0.9965  \\
$A_{R}/A_{5500 \text{\AA}}$ & 0.8187 & 0.7854 & 0.7683 \\
$A_{I}/A_{5500 \text{\AA}}$ &  0.5963  & 0.5641  & 0.5319  \\  \hline
$A_{J}/A_{5500 \text{\AA}}$ & 0.2981  & 0.2694  & 0.2380  \\
$A_{H}/A_{5500 \text{\AA}}$ & 0.1825 & 0.1716 & 0.1410 \\
$A_{Ks}/A_{5500 \text{\AA}}$ &  0.1178  & 0.1160  & 0.0853  \\  \hline
$A_{F555W\,\rm{ACS}}/A_{5500 \text{\AA}}$ & 1.0367 & 1.0423 &1.0356  \\
$A_{F606W,\rm{ACS}}/A_{5500 \text{\AA}}$ & 0.9339 & 0.9178  & 0.9058   \\
$A_{F814W,\rm{ACS}}/A_{5500 \text{\AA}}$ & 0.5699 & 0.5406  & 0.5079  \\  \hline
$A_{F555W,\rm{WFC3}}/A_{5500 \text{\AA}}$ & 1.0482 & 1.0521  & 1.0414  \\
$A_{F606W,\rm{WFC3}}/A_{5500 \text{\AA}}$ & 0.9304 & 0.9138 & 0.9019  \\
$A_{F814W,\rm{WFC3}}/A_{5500 \text{\AA}}$ & 0.6001  & 0.5671 & 0.5354  \\   
$A_{F110W,\rm{WFC3}}/A_{5500 \text{\AA}}$ & 0.3379 & 0.3184  &  0.2855 \\
$A_{F125W,\rm{WFC3}}/A_{5500 \text{\AA}}$ & 0.2881 & 0.2686 &  0.2372 \\   
$A_{F160W,\rm{WFC3}}/A_{5500 \text{\AA}}$ &0.2042  & 0.1907  &  0.1603 \\    \hline
\end{tabular}
\label{table:ExtinctionRatios} 
\end{table}

\begin{table}
\caption{\large Predicted extinction coefficients for the Landolt, 2MASS, ACS, and WFC3 filter transmission curves convolved with the synthetic spectrum of a $T_{\rm{eff}}=6,750$ Kelvin, [Fe/H]$=0$, $\log{g}=1.0$ star as well as 0.40 mag of extinction at $\lambda=5500$\AA.} 
\large
\centering 
\begin{tabular}{|l|c|c|c|} 
\hline\hline\hline 
 % & \rot{  \citet{1989ApJ...345..245C}  }  &   \rot{\citet{1999PASP..111...63F} } &  \rot{\citet{2007ApJ...663..320F}}    \\  
    & CCM89  &  F99 &  FM07    \\  
 \hline\hline
$A_{I}/E(V-I)$ & 1.4524 & 1.2721 & 1.1304 \\
$A_{U}/A_{5500 \text{\AA}}$ & 1.5395 & 1.5407  & 1.5151 \\
$A_{B}/A_{5500 \text{\AA}}$ & 1.3129  & 1.3233 & 1.2979  \\
$A_{V}/A_{5500 \text{\AA}}$ & 1.0078  & 1.0085  & 1.0035  \\
$A_{R}/A_{5500 \text{\AA}}$ & 0.8258 & 0.7933 & 0.7768 \\
$A_{I}/A_{5500 \text{\AA}}$ &  0.5969  & 0.5646  & 0.5325  \\  \hline
$A_{J}/A_{5500 \text{\AA}}$ & 0.2900  & 0.2702  & 0.2388  \\
$A_{H}/A_{5500 \text{\AA}}$ & 0.1826 & 0.1717 & 0.1411 \\
$A_{Ks}/A_{5500 \text{\AA}}$ &  0.1178  & 0.1160  & 0.0853  \\  \hline
$A_{F555W\,\rm{ACS}}/A_{5500 \text{\AA}}$ & 1.0435 & 1.0503 &1.0432  \\
$A_{F606W,\rm{ACS}}/A_{5500 \text{\AA}}$ & 0.9491 & 0.9360  & 0.9244   \\
$A_{F814W,\rm{ACS}}/A_{5500 \text{\AA}}$ & 0.5711 & 0.5417  & 0.5090  \\  \hline
$A_{F555W,\rm{WFC3}}/A_{5500 \text{\AA}}$ & 1.0636 & 1.0698  & 1.0583  \\
$A_{F606W,\rm{WFC3}}/A_{5500 \text{\AA}}$ & 0.9447 & 0.9309 & 0.9195  \\
$A_{F814W,\rm{WFC3}}/A_{5500 \text{\AA}}$ & 0.6016  & 0.5684 & 0.5368  \\   
$A_{F110W,\rm{WFC3}}/A_{5500 \text{\AA}}$ & 0.3429 & 0.3234  &  0.2904 \\
$A_{F125W,\rm{WFC3}}/A_{5500 \text{\AA}}$ & 0.2895 & 0.2699 &  0.2385 \\   
$A_{F160W,\rm{WFC3}}/A_{5500 \text{\AA}}$ &0.2045  & 0.1910  &  0.1606 \\    \hline
\end{tabular}
\label{table:ExtinctionRatios2} 
\end{table}

I then follow the sample selection outlined in Section 4 and Section 5 of \citet{2011ApJ...733..124S}, which is based off the work of \citet{2006ApJ...652.1133M}, including the same period-luminosity relations from \citet{1999AcA....49..223U}, and sample-cuts based on colour and variability amplitude ratios meant to reduce blends. However, I make a few modifications to the methodology:
\begin{enumerate}
 \item I use recursive 3-$\sigma$ outlier rejection rather than iterative weighted-median sigma clipping to remove outliers.  This consistently leads to 3 of the 231 clean Cepheids being removed from the fit, independent of the assumptions made in the fit. 
 \item The errors in the relations are computed assuming homoscedastic errors on the data points. I argue that the $\chi^{2}$ estimates from the method of \citet{2006ApJ...652.1133M} and \citet{2011ApJ...733..124S} are problematic.  The various sources of systematic error (such as the uncertain extinction curve) are in need of characterisation. and these will add to the existing errors in unexpected ways, rather than simply as a re-scaling of existing errors. Further, there is no correlation given for the errors on various variables, such as the error on the mean $I$-band magnitude or $V$-band magnitude of the Cepheids, which could be correlated. Thus it is actually impossible to compute a $\chi^{2}$ from the available information even if one assumes that all sources of error are known, since $\chi^{2}$ is a correlation-dependent quantity. 
  \item I fit for the metallicity-dependence of the dereddened Cepheid luminosities concurrently with the fit for the relative distance modulus of M101 with respect to the LMC using a multilinear least-squares fit. In contrast, \citet{2011ApJ...733..124S} computed the metallicity-dependence using two sequential least-squares fit, with the distance modulus computed first independently of the metallicity-dependence, and the scatter to the first relation then regressed with respect to the metallicity. 
  \item \citet{2011ApJ...733..124S} set their metallicity zero-point to [O/H]$=8.50$, which is thus their assumed metallicity for the LMC. I instead use [O/H]$=8.36$, as argued by \citet{2011ApJ...729...56B}, which was assumed by \citet{2011ApJ...730..119R} and is consistent with the measurement of [O/H]$=8.37$ in HII regions and supernovae remnants of the LMC  \citep{1990ApJS...74...93R}.
  % \citet{2008A&A...488..731R} measured a mean abundance of [Fe/H]$=-0.33$ for 22 LMC Cepheids, from which we can estimate [O/H]$=-0.48$ by assuming [O/Fe]$=-0.15$, from the spectroscopic study of 5 OB-type main-sequence LMC stars of  \citet{2002A&A...396...53R}.  The zero points of the stellar abundances can be set by the solar abundance of $12+\log$[O/H]$=8.69$ from \citet{2009ARA&A..47..481A}. The mean of these two estimates yields [O/H]$_{\rm{LMC}}=8.29$, which I adopt as a zero-point. In contrast, the expected mean metallicity for the sample of M101 Cepheids is [O/H]$=8.56$  \citep{2007ApJ...656..186B}. 
    \item As this investigation is concerned with quantifying one particular systematic error, the estimates of various parameters state only the maximum-likelihoood values and the statistical errors. In contrast to \citet{2011ApJ...733..124S} who report estimates of the total systematic error separate from their statistical errors. 
\end{enumerate}

\section{Results}
\label{sec:Results}

\subsection{Metallicity-Dependence as a Free Parameter}
\label{sec:FreeParam}
Assuming the extinction curve of \citet{1989ApJ...345..245C}, I obtain:
\begin{equation}
\begin{split}
  {\Delta}{\mu}_{\rm{LMC}} = (10.663\pm0.031) + (-0.433\pm0.145)(\rm{[O/H]}-8.36)   \\
\delta = 0.1680, \\
\end{split}
\label{EQ:CCM89_MD}
\end{equation}
where $\delta$ is the 1-$\sigma$ scatter on the relation.  If I assume the extinction curve of \citet{1999PASP..111...63F}:
\begin{equation}
\begin{split}
  {\Delta}{\mu}_{\rm{LMC}} = (10.692\pm0.031) + (-0.406\pm0.144)(\rm{[O/H]}-8.36)   \\
\delta = 0.1671, \\
\end{split}
\label{EQ:F99_MD}
\end{equation}
and for the extinction curve of \citet{2007ApJ...663..320F}:
\begin{equation}
\begin{split}
  {\Delta}{\mu}_{\rm{LMC}} = (10.716\pm0.031) + (-0.384\pm0.144)(\rm{[O/H]}-8.36)   \\
\delta = 0.1674, \\
\end{split}
\label{EQ:FM07_MD}
\end{equation}
I reiterate that the errors quoted are random errors resulting from the assumption of  homoscedastic errors on the data points. A complete inventory of the systematics involved in Cepheid determinations and their associated correlations is not available at this time and is beyond the scope of this work. Only one systematic is being focused on: the  shape of the extinction curve. 

The relative distance estimate from Equations 2,3,4 is $\sim$0.03 mag greater than that from  \citet{2011ApJ...733..124S} in spite of assuming the same extinction curve and nearly the same selection function. The offset is largely due to the downward adjustment of the assumed LMC metallicity by $\sim$0.14 dex. 

For the distance relations fit by Equation \ref{EQ:CCM89_MD} through to Equation \ref{EQ:F99_MDfree} in the following subsection, the same three outliers and no others are always removed: F2-3598, F1-3546, and F2-1458. Their respective inferred distances are $\sim$0.60 mag, $\sim$-0.70 mag, and $\sim$1.20 mag further than the relations. Due to these large offsets and the fact that it's the same three outliers removed each time, I argue that the recursive 3-$\sigma$ outlier rejection does not detrimentally bias the results. 

The metallicity dependence derived, $\gamma = dM_{I}/d[M/H] \approx 0.40$ mag dex$^{-1}$, is substantially lower than the values of 0.72 mag dex$^{-1}$ and 0.80 mag dex$^{-1}$ that  \citet{2011ApJ...733..124S} derive using two different methods. That may be due to the different method of fitting. The value derived here is closer to other literature estimates derived from other estimates, as will be shown in the subsequent subsection. 

\subsection{Metallicity-Dependence Fixed to Literature Values}
It is of interest to verify how results differ when the metallicity-dependence is fixed to literature values. \citet{2004ApJ...608...42S}  estimated $\gamma = -0.24\pm0.05$ mag dex$^{-1}$ by comparing tip of the red giant branch magnitudes to Cepheid magnitudes for 7 nearby galaxies images in $V$ and $I$ with HST, as well as 10 additional values found in the literature. \citet{2011A&A...534A..95S} estimated $\gamma = -0.23\pm0.10$ mag dex$^{-1}$ for a sample of 51 Cepheids from the SMC, the LMC, and the Milky Way. I adopt the value of $\gamma = -0.23$ mag dex$^{-1}$ though the two values are so similar that the effect of this choice is negligible. 

The relation obtained for the extinction curve of \citet{1989ApJ...345..245C} is:
\begin{equation}
\begin{split}
  {\Delta}{\mu}_{\rm{LMC}} = (10.622\pm0.011)   \\
\delta = 0.1687, \\
\end{split}
\label{EQ:CCM89_MDfree}
\end{equation}
for the extinction curve of \citet{1999PASP..111...63F}:
\begin{equation}
\begin{split}
  {\Delta}{\mu}_{\rm{LMC}} = (10.657\pm0.011)  \\
\delta = 0.1676, \\
\end{split}
\label{EQ:F99_MDfree}
\end{equation}
and for the extinction curve of \citet{2007ApJ...663..320F}:
\begin{equation}
\begin{split}
  {\Delta}{\mu}_{\rm{LMC}} = (10.685\pm0.011)  \\
\delta = 0.1678, \\
\end{split}
\label{EQ:FM07_MDfree}
\end{equation}

M101 is shifted to a distance that is $\sim$0.04 mag further away in all three cases relative to the values obtained when the metallicity-dependence is floated as a free parameter. That offset comes from the product of the mean [O/H] of M101 relative to the LMC ($\sim$0.20 dex) and the shift in the slope of the metallicity-dependence ($\sim$0.15-0.20 mag dex$^{-1}$). 

Though this was interesting as an exercise,  I note that this assumption of a metallicity-dependence form the literature is inconsistent with other assumptions in this investigation. \citet{2011A&A...534A..95S} assumed a universal, constant extinction function of $R_{V}=3.23$ from \citet{1989ApJ...345..245C}, corresponding to $A_{I}/E(V-I) \approx 1.50$. That's marginally shallower than the LMC value of $A_{I}/E(V-I) = 1.44 \pm 0.03$ \citep{2003ApJ...590..284U}, and substantially shallower than the Milky Way value of $A_{I}/E(V-I) \approx 1.27$ suggested by the work of \citet{2011ApJ...737..103S}. It may also be distinct from the SMC value. There is an urgent need for an investigation of the metallicity-dependence of Cepheid luminosities that accounts for extinction curve uncertainties and variations in the Milky Way Galaxy and its satellites. 

\subsection{Shifting the Maximum-Colour Cut}
 \citet{2011ApJ...733..124S} removed Cepheids with $E(V-I) > 0.39$ from their fit, on the basis that they might either be red blends, or have high-reddening values that would lead to an error linearly proportional to the error in the assumed extinction curve. Of interest would be to test which of the three extinction curves yields a distance least sensitive to the choice of $E(V-I)$ cut (hereafter: $E(V-I)_{\rm{Max}}$). For this calculation, I fix the metallicity-dependence to the literature value of  $\gamma = -0.23$ mag dex$^{-1}$ \citep{2011A&A...534A..95S} to restrict the number of degrees of freedom, though it is noted that the results are nearly identical if I fix the metallicity dependence to the values found in Equations  \ref{EQ:CCM89_MD},  \ref{EQ:F99_MD}, and \ref{EQ:FM07_MD}.

\begin{figure}
\begin{center}
\includegraphics[totalheight=0.32\textheight]{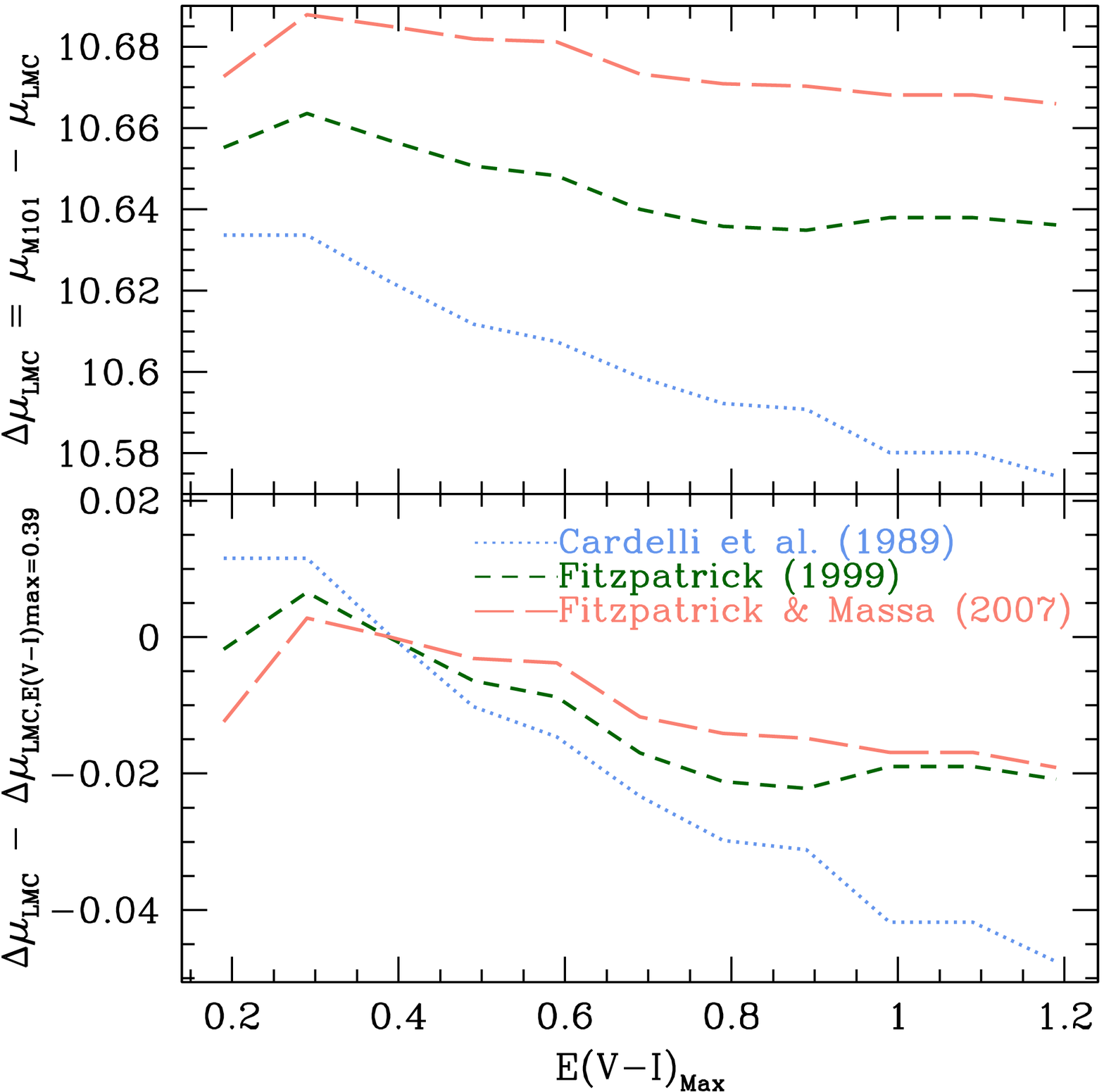}
\end{center}
\caption{\large TOP: The distance modulus of M101 relative to that of the LMC as a function of the extinction curves and as the maximum reddening $E(V-I)_{\rm{Max}}$ of Cepheids included in the calculation. The extinction curve of  \citet{2007ApJ...663..320F} is the least sensitive to the choice of $E(V-I)_{\rm{Max}}$, whereas that of  \citet{1989ApJ...345..245C} is the most sensitive. BOTTOM: Same as top panel, but with the distance modulus for  $E(V-I)_{\rm{Max}}=0.39$ (the value assumed by  \citealt{2011ApJ...733..124S}), shifted to zero.} 
\label{Fig:ReddeningCutEffect}
\end{figure}

 With this test, of which I plot the results in Figure \ref{Fig:ReddeningCutEffect}, the extinction curve of \citet{2007ApJ...663..320F} is the most successful at matching the data for M101, whereas that of \citet{1989ApJ...345..245C} which is the most standard in the Cepheid literature is the least successful. The inferred relative distance modulus decreases rapidly as more reddened sources are included when I assume the extinction curve of \citet{1989ApJ...345..245C}, exactly what one would expect if the extinction curve is too shallow. The distance modulus also decreases when I assume the extinction curves of \citet{1999PASP..111...63F} \citet{2007ApJ...663..320F}, but by a much smaller amount, $\sim -0.027$ and $\sim -0.017$ respective decreases in distance modulus per 1 magnitude increase in $E(V-I)_{\rm{Max}}$. In contrast, use of the extinction curve of \citet{1989ApJ...345..245C} yields a much steeper sensitivity of $\sim -0.063$. These results are listed in Table \ref{table:ColourCutOffset} .
 
 \begin{table}
\caption{\large The slope of the shift in inferred distance modulus per 1 magnitude increase in the reddening cut, as a function of assumed extinction curve. The curve of \citet{2007ApJ...663..320F} yields the smallest dependence to the choice of colour cut. } 
\large
\centering 
\begin{tabular}{|l|c|} 
\hline\hline\hline 
CCM89  &  $-$0.063 \\
F99 &   $-$0.027 \\
FM07 &   $-$0.017 \\  
\hline\hline 
\end{tabular}
\label{table:ColourCutOffset} 
\end{table}
 
\begin{table}
\caption{\large Scatter in distance modulus as a function of colour-cut on the Cepheid sample and the choice of the extinction curve. The extinction curve of \citet{2007ApJ...663..320F} yields the smallest increase in scatter as the colour cut is increased. } 
\large
\centering 
\begin{tabular}{|c|ccc|} 
\hline\hline\hline 
$E(V-I)_{\rm{Max}}$    & CCM89  &  F99 &  FM07    \\  
 \hline\hline
0.39 & 0.1687 & 0.1676 & 0.1678 \\ 
0.99 & 0.1844 & 0.1839 & 0.1766 \\ 
\hline\hline 
\end{tabular}
\label{table:DistanceScatter} 
\end{table}

The same conclusions are reached if one would rather track the scatter in the relations rather than the shift in distance modulus, for which the results are listed in Table \ref{table:DistanceScatter}. Assuming the extinction curve of  \citet{1989ApJ...345..245C}, I find that the scatter increases from $\sigma=0.1687$ at $E(V-I)_{\rm{Max}} = 0.39$ to $\sigma=0.1844$ at $E(V-I)_{\rm{Max}} = 0.99$. With the extinction curve of \citet{1999PASP..111...63F}, the scatter increases from $\sigma=0.1676$ to  $\sigma=0.1839$. Finally, with the extinction curve of \citet{2007ApJ...663..320F}, the scatter increases from $\sigma=0.1678$ to  $\sigma=0.1766$. The increase in variance for the shallowest extinction curve as the reddening cut is increased is only $\sim$10\%, less than half what it is for the other two extinction curves.  

\section{Discussion and Conclusion}
In this investigation I have demonstrated that the Cepheid-distance to the spiral galaxy M101 is measured with a reduced scatter and reduced sensitivity to systematics if I assume the mean Galactic extinction curve of \citet{2007ApJ...663..320F}, which is steeper than the canonical $R_{V}=3.1$ extinction curve from \citet{1989ApJ...345..245C}.  The steeper extinction curve increases the distance modulus to M101 by $\sim$0.06 mag. The combination of the different extinction curve, and the downward adjustment in the assumed metallicity of the LMC, yields a distance modulus estimate that is $\sim$0.09 mag greater than that of  \citet{2011ApJ...733..124S}, to $ {\Delta}\mu_{\rm{LMC}} \approx 10.72 \pm 0.03$. The best-fit metallicity-dependence is $dM_{I}/d\rm{[O/H]} \approx (-0.38 \pm 0.14)$ mag dex$^{-1}$. 

The mean Galactic extinction curve of \citet{2007ApJ...663..320F} yields a superior fit to the M101 Cepheids, but it should not be expected to apply in general. \citep{2013ApJ...775L..16K}  have shown that the attenuation curves of galaxies correlates with their spectral type and specific-star-formation rate. Meanwhile, it's by now well-documented that the shape of extinction curve is a function of location even in one particular galaxy \citep{2009ApJ...707..510Z,2013ApJ...779...38P,2014ApJ...785..136D}. I thus argue that a new phenomenological model to deal with interstellar extinction is needed, both to understand extragalactic stellar populations and to precisely and accurately calibrate the distance ladder. 

The referee brought up the possibility of varying $R_{V}$. Given that the investigation of this work had access to data in only two bandpasses, varying $R_{V}$ from a fixed family of extinction curves is indistinguishable from varying the extinction curve at fixed $R_{V}$. From the formalism of \citet{1989ApJ...345..245C}, one obtains $A_{I}/E(V-I) = 1.145$ at $R_{V} \approx 2.33$, a value common in Type Ia SNe cosmology \citep{2013ApJ...779...38P}. The distinction cannot be visited in this work, it could in principle be investigated if there were a larger number of bandpasses available, in particular $B$-band would obviously be a significant degeneracy-breaker. 

Some readers may worry that the era of worrying about extinction as a source of error in the distance ladder is ending. For example,  \citet{2012ApJ...745..156R} calibrated the near-IR period-luminosity relations for a sample of 68 Cepheids in the Andromeda galaxy (M31). This strategy is promising,  as the extinction in WFC3 filter $F160W$ is only $\sim$30\% that of the extinction in the Landolt $I$-band filter, and the discrepancy in the predicted value of $A_{F160W}$ between different extinction curves is only $\sim$5\% of the size of $A_{5500\text{\AA}}$ (see Table \ref{table:ExtinctionRatios}).  However, it will also be harder to measure the Hubble flow to longer distances given the $\sim$300\% reduction in resolution at the longer wavelength, in addition to the possibility that Cepheids may be further blended at these longer wavelengths, for example by AGB stars which are intrinsically bright, red, and numerous. Further, regardless of this issue, the existing trove of optical Cepheid data to nearby galaxies yields a probe of the interstellar medium that is distinct from others currently in use.

\section*{Acknowledgments}
I thank Benjamin J. Shappee, Brian P. Schmidt, and Martin Asplund, for helpful discussions. I thank the anonymous referee for a constructive review of the manuscript. \\
DMN was  supported by the Australian Research Council grant FL110100012.

\bibliography{Nataf_M101Cepheids_MNRASv2}
%\begin{thebibliography}{42}
%\end{thebibliography}

\end{document}